\documentclass[11pt]{article}
\usepackage[T1]{fontenc}
\usepackage{lmodern}
\usepackage[utf8]{inputenc}
\usepackage[margin=1in]{geometry}
\usepackage{amsmath,amssymb,graphicx,array,booktabs,longtable,calc}
\usepackage[hidelinks]{hyperref}
\title{Temporal Seam Score for Assessing Continuity at Known Transitions in Time Series}
\author{Hongxiao Jin\\[4pt]\small Department of Earth and Environmental Sciences, Lund University, Lund, Sweden}
\date{}
\begin{document}
\maketitle
\begin{abstract}
Long environmental records combine observations from successive observing or processing systems, creating temporal seams that may reflect imperfect harmonization or environmental change. We introduce the Temporal Seam Score (S-score), a signed, robust diagnostic of continuity at a known transition. A seam-blind model accounts for trend, seasonality, and optional covariates. The residual contrast between symmetric windows around the transition is standardized against matched within-segment reference contrasts using their median and median absolute deviation. The sign describes departure relative to the reference center. In simulations, a 20\% shift produced median absolute scores of 2.58 at a residual coefficient of variation of 0.10 and 0.84 at 0.50. In a MODIS-to-Sentinel-3 vegetation record, 112 of 500 sites yielded finite scores, and 89 met all reliability criteria. Among these sites, median S was \ensuremath{-}0.020 and median absolute S was 0.717, indicating little systematic departure from the reference centers. The S-score supports continuity assessment without temporal overlap, with explicit diagnostics of data sufficiency and reference support. It complements harmonization and attribution analyses but does not establish practical equivalence or identify the cause of a discontinuity.

\end{abstract}
\textbf{Keywords} Harmonization; Environmental monitoring; Remote sensing; Robust standardization; Quality assurance

\hypertarget{introduction}{%
\section{1 Introduction}\label{introduction}}

Long environmental records increasingly combine observations from successive generations of observing or processing systems. Sensors are replaced, station instruments are upgraded, retrieval algorithms are revised, and reanalysis and model products move between production generations. Although these transitions are often documented in metadata, the resulting records are commonly analyzed as continuous time series. Even a modest incompatibility at one transition can affect estimated trends, extremes, phenological metrics, disturbance histories, and treatment effects.

Conventional validation metrics address only part of this problem. Bias, root mean square error, regression slope, and Lin's concordance correlation coefficient compare paired observations when two systems operate concurrently (Lin, 1989). Such overlap is central to cross-sensor calibration and harmonization, as illustrated by the Harmonized Landsat-Sentinel-2 framework (Claverie et al., 2018). Many composite records, however, contain little or no overlap. For satellite missions, overlap may coincide with degradation of the older sensor or early operational instability of the newer sensor. Paired agreement may therefore be uncertain and does not guarantee temporal continuity after temporal aggregation, quality screening, gap handling, or downstream retrieval. Time-series homogenization methods can detect and correct inhomogeneities using reference records (Venema et al., 2012), but an auditable diagnostic is still needed to quantify continuity at a transition in the final composite record.

Established methods detect time-series shifts in distribution or regression structure. These approaches include cumulative-sum procedures (Page, 1954), structural-break tests (Chow, 1960), intervention analysis (Box \& Tiao, 1975), and methods for identifying single or multiple change points (Bai \& Perron, 1998; Pettitt, 1979). Remote-sensing methods such as BFAST (Verbesselt et al., 2010), LandTrendr (Kennedy et al., 2010), CCDC (Zhu \& Woodcock, 2014), and DBEST (Jamali et al., 2015) identify disturbances or changes in trend and seasonality from spectral trajectories. These approaches are suited to locating unknown changes or characterizing the changes themselves. Continuity assessment poses a distinct question: given a known transition time, how exceptional is the junction relative to ordinary temporal transitions within the contributing segments?

We introduce the Temporal Seam Score (S-score, or \emph{S}), a signed, robustly standardized measure designed to answer this question. After trend, seasonality, and optional covariate effects are represented by a seam-blind model, the residual contrast across a symmetric window around the transition is standardized against matched reference contrasts wholly within the contributing segments. The S-score is defined as the standardized departure of the observed seam contrast from ordinary within-segment reference contrasts. The sign of \emph{S} indicates the direction of departure relative to the reference center, while \textbar{}\emph{S}\textbar{} describes how exceptional it is relative to within-segment variability.

We develop the S-score as a general method. Simulation is conducted to quantify how \textbar{}\emph{S}\textbar{} varies with imposed seam magnitude shift and residual variability. A MODIS-to-Sentinel-3 vegetation record demonstrates implementation, interpretation, and reliability screening of S-score. The broader framework applies to known transitions among instruments, station configurations, processing algorithms, reanalyses, and other generations of environmental data systems.

\hypertarget{temporal-seams-and-continuity-assessment}{%
\section{2 Temporal seams and continuity assessment}\label{temporal-seams-and-continuity-assessment}}

\hypertarget{definition-of-a-temporal-seam}{%
\subsection{2.1 Definition of a temporal seam}\label{definition-of-a-temporal-seam}}

Let a composite record contain two consecutive segments A and B separated by a known transition time \ensuremath{\tau}\textsubscript{0}:

\begin{equation}
y_{t} = \left\{ \begin{matrix}
y_{t}^{(A)}, & t \leq \tau_{0}, \\
y_{t}^{(B)}, & t > \tau_{0}.
\end{matrix} \right.\ \tag{1}
\end{equation}

The segment labels A and B identify a change in the observing or processing system. They do not imply that the underlying environmental process changed at \ensuremath{\tau}\textsubscript{0}. Records containing several known seams can be analyzed by evaluating each transition separately or through a hierarchical framework.

In this study, temporal continuity means that the residual contrast across a known transition is consistent with ordinary within-segment temporal variability. Three questions should be distinguished. Detection asks whether an unknown change occurred and where it occurred. Continuity assessment asks how exceptional a known seam is relative to matched within-segment transitions. Attribution asks why a discontinuity occurred. The S-score addresses continuity assessment. A large absolute S-score can motivate attribution analysis using information about instrument changes, environmental disturbances, land-cover change, management, or other events near the seam.

\hypertarget{seam-blind-reference-model}{%
\subsection{2.2 Seam-blind reference model}\label{seam-blind-reference-model}}

Expected temporal behavior is modeled without a segment indicator, transition step, or sensor-specific intercept:

\begin{equation}
y_{t} = f\left( t,\mathbf{x}_{t} \right) + \varepsilon_{t}\tag{2}
\end{equation}

Here, \emph{y\textsubscript{t}} is the observed quantity, \emph{t} is time, \emph{\textbf{x}\textsubscript{t}} contains optional explanatory variables, \emph{f} is the expected-value model, and \emph{\ensuremath{\varepsilon}\textsubscript{t}} is the residual. The model may contain a trend, seasonality, meteorological covariates, or other specific explanatory terms. A generalized additive model with a seasonal spline and a conservative smooth trend is appropriate for many dense environmental records, including greenness time series (Cai et al., 2017). Annual records, such as annual phenology metrics, may require only a gradual trend (Jin et al., 2019). The model should capture expected temporal variation while preserving a local discontinuity at \emph{\ensuremath{\tau}\textsubscript{0}}.

\hypertarget{observed-seam-contrast-with-aggregated-residuals}{%
\subsection{2.3 Observed seam contrast with aggregated residuals}\label{observed-seam-contrast-with-aggregated-residuals}}

Residuals are aggregated to a temporal unit \emph{k} appropriate for application:

\begin{equation}
R_{k} = \mathcal{A}\left\{ {\widehat{\varepsilon}}_{t}:t \in k \right\}\tag{3}
\end{equation}

The aggregation operator \(\mathcal{A}\) may be a mean, median, seasonal integral, phenological metric, or another prespecified summary. The aggregation must use comparable groups on both sides of the seam. Missing observations can be addressed through matched dates, common temporal bins, model-based weights, or an explicit coverage rule.

For a symmetric window containing \emph{W} aggregated periods on each side, the signed seam contrast is:

\begin{equation}
\mathrm{\Delta}_{seam}^{(W)} = \frac{1}{W}\sum_{j = 1}^{W}R_{\tau_{0} + j} - \frac{1}{W}\sum_{j = 0}^{W - 1}R_{\tau_{0} - j}\tag{4}
\end{equation}

Positive values of \(\Delta_{\mathrm{seam}}\) indicate an upward residual contrast and negative values indicate a downward residual contrast after accounting for the fitted reference model. Short windows emphasize local discontinuity but are sensitive to individual extremes. Longer windows reduce sampling variability but may combine the seam with gradual environmental change. A proper window should therefore reflect the process timescale, preferably with a window-sensitivity analysis.

\hypertarget{matched-within-segment-reference-transition}{%
\subsection{2.4 Matched within-segment reference transition}\label{matched-within-segment-reference-transition}}

Ordinary temporal variability must be estimated on the same scale as the observed seam contrast. For each eligible pseudo-transition \emph{\ensuremath{\tau}} in the set \(\mathcal{P}\), an identical symmetric contrast is calculated:

\begin{equation}
\mathrm{\Delta}_{\tau}^{(W)} = \frac{1}{W}\sum_{j = 1}^{W}R_{\tau + j} - \frac{1}{W}\sum_{j = 0}^{W - 1}R_{\tau - j},\ \tau \in \mathcal{P}.\tag{5}
\end{equation}

If a gap period is excluded between the two windows used to calculate the seam contrast, a gap of the same duration must also separate the windows used to calculate each within-segment reference contrast. Both windows before and after \(\tau\) of Eq. (5) must lie wholly within the same record segment, and contrasts crossing the observed seam are excluded, so as to compare the observed reference contrasts over identical windows.

The reference center and statistical spread are robustly estimated as:

\begin{equation}
{\widetilde{\mathrm{\Delta}}}_{ref} = {median\ }_{\tau \in \mathcal{P}}\left\lbrack \mathrm{\Delta}_{\tau}^{(W)} \right\rbrack\ ,{\widehat{\sigma}}_{ref} = 1.4826 \times {MAD\ }_{\tau \in \mathcal{P}}\left\lbrack \mathrm{\Delta}_{\tau}^{(W)} \right\rbrack\ .\tag{6}
\end{equation}

The reference center is the median contrast, and the reference spread is computed as 1.4826 times the median absolute deviation (MAD). These robust estimates limit the influence of occasional genuine disturbances among the reference transitions.

\hypertarget{temporal-seam-score}{%
\subsection{2.5 Temporal seam score}\label{temporal-seam-score}}

The temporal seam score (S-score), denoted \emph{S}, is defined as:

\begin{equation}
S = \frac{\mathrm{\Delta}_{seam}^{(W)} - {\widetilde{\mathrm{\Delta}}}_{ref}}{{\widehat{\sigma}}_{ref}}.\tag{7}
\end{equation}

An S-score near zero indicates that the observed seam contrast lies near the center of the within-segment reference distribution. Its sign indicates the direction of departure relative to the reference center: S \textgreater{} 0 means that \(\Delta_{\mathrm{seam}}\) lies above the reference median, whereas S \textless{} 0 means that it lies below that median. The sign of \(\Delta_{\mathrm{seam}}\), rather than S, gives the raw upward or downward direction of the residual contrast. The signs of S and \(\Delta_{\mathrm{seam}}\) can differ when the reference median is nonzero. The absolute magnitude \textbar S\textbar{} indicates how exceptional the departure is relative to ordinary within-segment transitions.

The S-score is undefined when the estimated reference spread is zero. Because contrasts calculated from overlapping windows are dependent, their nominal number overstates the independent information available. Fewer than ten valid reference contrasts, or the absence of reference contrasts from either segment, should trigger a limited-reference warning. This threshold represents a practical minimum rather than ten independent observations. A shorter \(W\) may provide more reference contrasts, but it produces a more local and generally noisier continuity assessment. In spatially replicated applications, a shared or hierarchical reference distribution may be estimated from comparable sites or pixels, provided that the analysis accounts for spatial dependence.

As a secondary sensitivity diagnostic, let \(\mathcal{D}_{AD}\) contain all adjacent residual differences for which both periods belong to the same segment. The adjacent-difference score is:

\begin{equation}
\begin{gathered}S_{AD} = \frac{\Delta_{seam}^{(W)}}{SD\left( \mathcal{D}_{AD} \right)},\\ \mathcal{D}_{AD} = \left\{ R_{k} - R_{k - 1}:k\text{ and }k - 1\text{ belong to the same segment} \right\}.\end{gathered}\tag{8}
\end{equation}

Here \(SD(\mathcal{D}_{AD})\) is the standard deviation of the eligible adjacent differences, each representing the change from period \emph{k}-1 to period \emph{k}. This denominator measures variability in adjacent-period changes, whereas \(\Delta_{seam}^{(W)}\) compares means calculated over \emph{W}-period windows. Because the numerator and denominator operate at different temporal scales, \(S_{AD}\) is reported only as a supplementary diagnostic.

\hypertarget{simulation}{%
\section{3 Simulation}\label{simulation}}

\hypertarget{simulation-design}{%
\subsection{3.1 Simulation design}\label{simulation-design}}

The simulation quantified the relationship between the S-score, seam magnitude, and residual variability, with additional checks of its finite-sample behavior and directional interpretation. To evaluate the scoring step separately from estimation of the seam-blind model, simulated records were generated directly as aggregated residual summaries:

\begin{equation}
R_{k} = \left\{ \begin{matrix}
\varepsilon_{k}, & k \leq \tau_{0} \\
\varepsilon_{k} + q, & k > \tau_{0}
\end{matrix} \right.\ ,\ \varepsilon_{k}\sim N(0,{\sigma_{r}}^{2}),\tag{9}
\end{equation}

where \emph{k} denotes the aggregated period, \(\tau_0\) is the known transition time, \emph{q} is the imposed level shift. Each record contained two 20-period segments. Both \emph{q} and \(\sigma_{r}\) were expressed relative to a positive baseline mean \(\mu_{pre}\), representing the underlying observation series before the transition. Relative residual variability was defined as \({CV}_{r} = \sigma_{r}/\mu_{pre}\). Expressing both the shift and variability relative to the same baseline made the S-score independent of its absolute scale.

The seam contrast used the final three periods of the first segment and the first three periods of the second segment. Matched three-period-versus-three-period reference contrasts were calculated wholly within each segment and pooled. Their center and spread were estimated using the median and 1.4826 MAD, respectively.

Positive shifts from 0\% to 100\% of the pre-seam mean were evaluated at \({CV}_{r}\) values of 0.05, 0.10, 0.20, and 0.50. Negative shifts were additionally simulated at \({CV}_{r} = 0.10\) to assess direction and symmetry. Each combination contained 2,000 independently generated records.

\hypertarget{simulation-results}{%
\subsection{3.2 Simulation results}\label{simulation-results}}

Under continuity, the signed S-score was centered near zero. Across the four residual-variability levels, median S ranged from \ensuremath{-}0.03 to 0.02. The corresponding median \textbar S\textbar{} values ranged from 0.74 to 0.77 because random contrasts produce nonzero absolute scores even when no discontinuity is present. A nonzero \textbar S\textbar{} therefore represents ordinary sampling variability and does not alone indicate a seam discontinuity.

Positive and negative shifts produced approximately symmetric S-scores (Figure 1). At \(CV_r\) = 0.10, shifts of \ensuremath{\pm}10\% produced median S-scores of approximately \ensuremath{\pm}1.27, while shifts of \ensuremath{\pm}20\% produced median scores of approximately \ensuremath{\pm}2.5. The sign of S agreed with the imposed shift direction in approximately 87\%--88\% of simulations with a 10\% shift and about 99\% with a 20\% shift. This agreement is specific to the simulation design, whose within-segment reference contrasts are centered near zero; in general, the sign of S describes departure relative to the estimated reference center.

\begin{figure}[htbp]
\centering
\includegraphics[width=.77\linewidth]{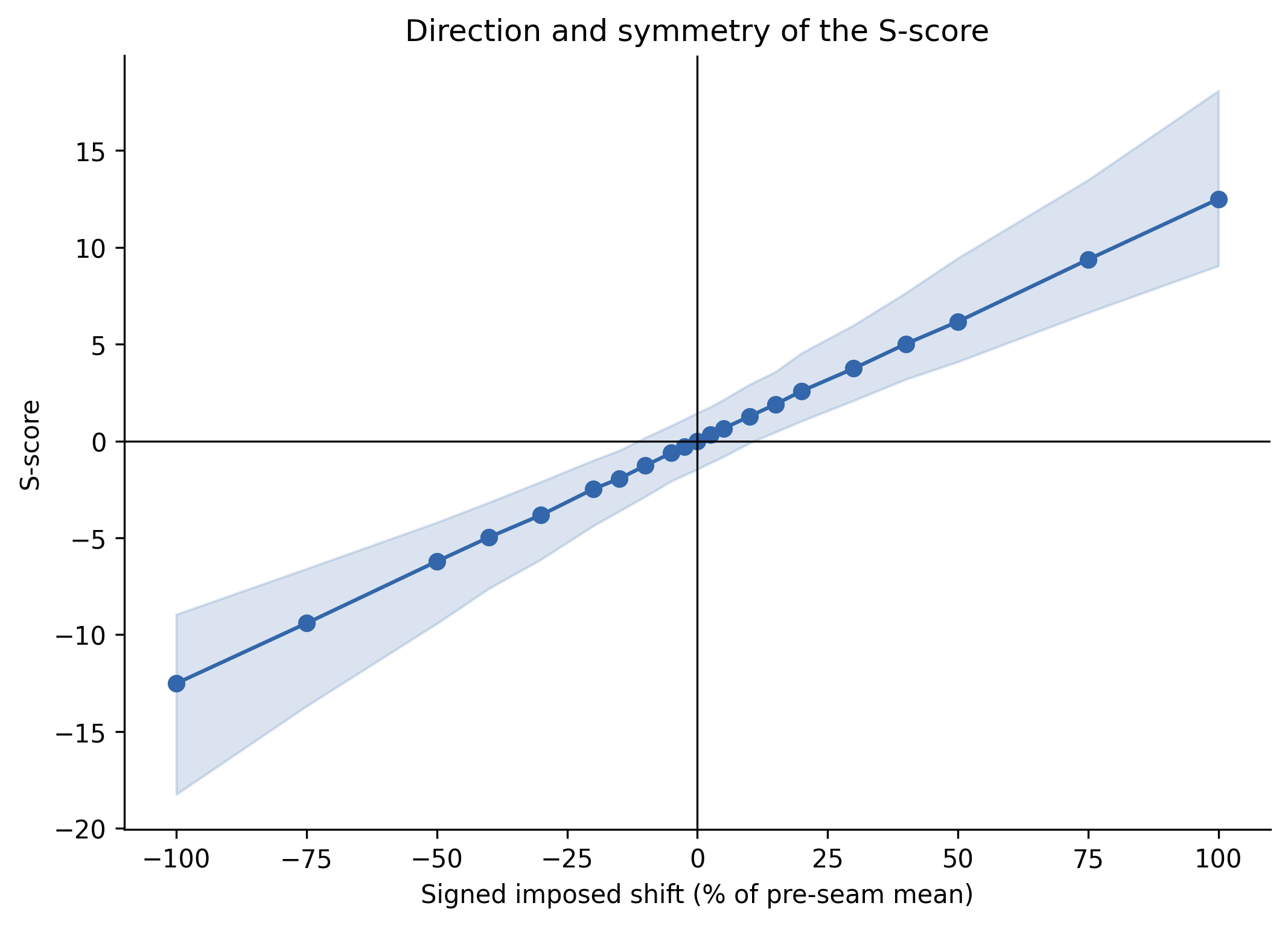}\par
\caption{\textbf{Direction and symmetry of the S-score.} Median signed S-score versus imposed positive and negative level shifts at \(CV_r\) = 0.10. Shading shows the 10th--90th percentile range across simulations. The sign of S describes departure relative to the estimated reference center; the imposed shift direction is known in this simulation.}
\end{figure}

The absolute S-score increased with seam magnitude shift but decreased as residual variability increased (Figure 2). For example, a 20\% shift produced median \textbar S\textbar{} values of 4.91, 2.58, 1.28, and 0.84 at \(CV_r\) values of 0.05, 0.10, 0.20, and 0.50, respectively. For a 50\% shift, the corresponding values were 12.63, 6.16, 3.12, and 1.31. At \(CV_r\) = 0.50, the median \textbar S\textbar{} for a 20\% shift was therefore only slightly greater than its value under continuity.

These results show that the S-score quantifies a seam relative to ordinary residual variability rather than measuring its absolute size. A given discontinuity becomes less distinguishable as residual variability increases. Consequently, a small \textbar S\textbar{} in a noisy record indicates limited ability to assess continuity, whereas a large \textbar S\textbar{} identifies a seam contrast that is exceptional relative to the variability of that record.

\begin{figure}[htbp]
\centering
\includegraphics[width=1\linewidth]{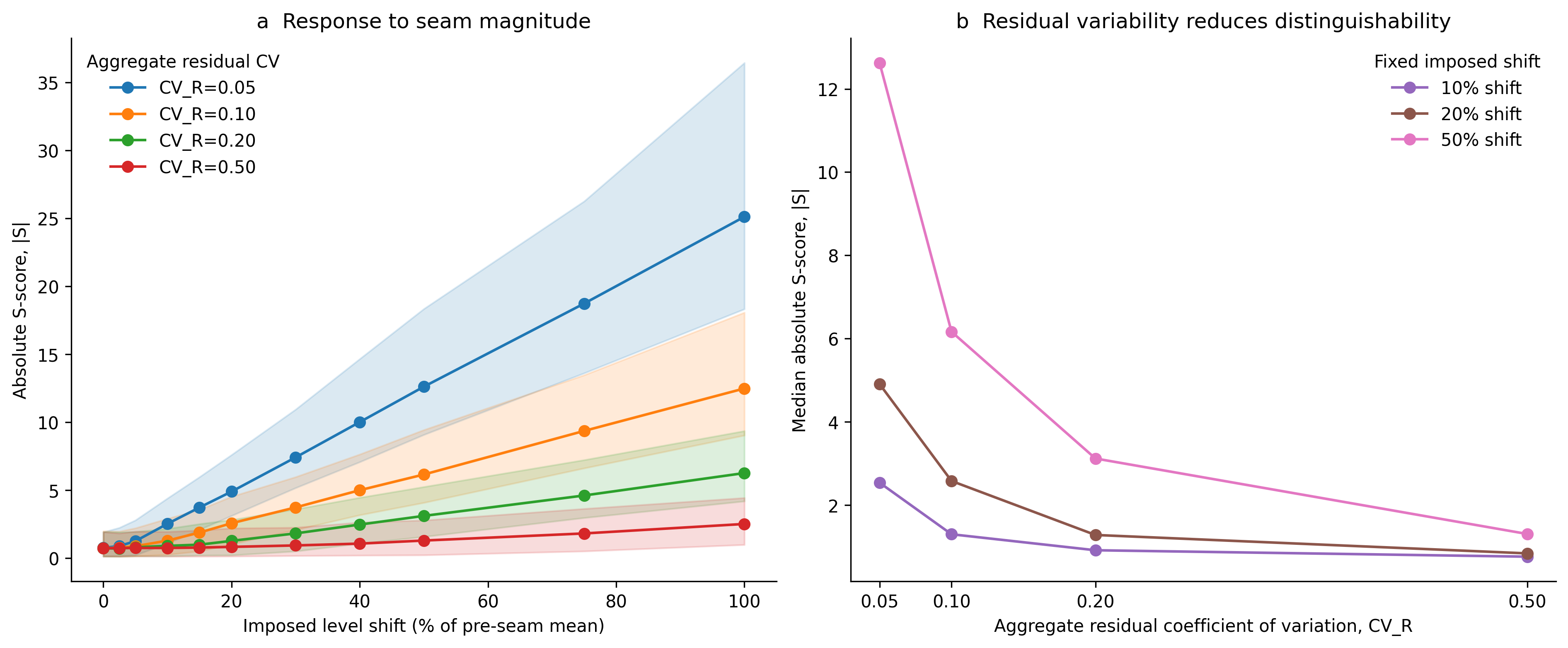}\par
\caption{\textbf{Response of the absolute S-score to seam magnitude and residual variability.} (a) Median \textbar S\textbar{} versus imposed positive level shift for selected aggregate residual coefficients of variation (\({CV}_{r}\)); shaded bands show the 10th--90th percentile range across 2,000 simulations. (b) Median \textbar S\textbar{} versus \({CV}_{r}\) for fixed level shifts. Each simulation used two 20-period segments, a symmetric three-period window, and pooled matched reference contrasts normalized by their median and 1.4826 MAD.}
\end{figure}

Based on standardized-score conventions, Table 1 provides descriptive categories of interpreting \textbar{}\emph{S}\textbar. Formal thresholds for these categories should account for segment length, window design, dependence, missingness, and the number and composition of reference contrasts. They are not equivalence criteria or error-controlled detection thresholds.

\begin{longtable}[]{@{}
  >{\raggedright\arraybackslash}p{(\columnwidth - 4\tabcolsep) * \real{0.1736}}
  >{\raggedright\arraybackslash}p{(\columnwidth - 4\tabcolsep) * \real{0.2065}}
  >{\raggedright\arraybackslash}p{(\columnwidth - 4\tabcolsep) * \real{0.6198}}@{}}
\caption{Rule-of-thumb interpretation of the absolute S-score.}\\
\toprule\noalign{}
\begin{minipage}[b]{\linewidth}\raggedright
Absolute S-score
\end{minipage} & \begin{minipage}[b]{\linewidth}\raggedright
Standardized seam magnitude
\end{minipage} & \begin{minipage}[b]{\linewidth}\raggedright
Descriptive interpretation
\end{minipage} \\
\midrule\noalign{}
\endhead
\bottomrule\noalign{}
\endlastfoot
\textbar S\textbar{} \textless{} 1 & Small & The seam contrast lies within one reference spread of the within-segment center. \\
1 \ensuremath{\leq} \textbar S\textbar{} \textless{} 2 & Moderate & The seam contrast lies between one and two reference spreads from the within-segment center. \\
2 \ensuremath{\leq} \textbar S\textbar{} \textless{} 3 & Large & The seam contrast lies between two and three reference spreads from the within-segment center and warrants further investigation. \\
\textbar S\textbar{} \ensuremath{\geq} 3 & Very large & The seam contrast lies at least three reference spreads from the within-segment center and warrants further investigation. \\
\end{longtable}

\hypertarget{demonstration-using-a-modis-to-sentinel-3-vegetation-record}{%
\section{4 Demonstration using a MODIS to Sentinel 3 vegetation record}\label{demonstration-using-a-modis-to-sentinel-3-vegetation-record}}

We demonstrated the S-score using a known transition from a historical MODIS archive to Sentinel-3 with the Plant Phenology Index (PPI, Jin \& Eklundh, 2014) record. The analysis assessed whether the temporal seam was exceptional relative to ordinary transitions within the two record segments before and after the transition on 1 January 2019. It was designed neither to locate an unknown change nor to attribute the seam to sensor replacement or an environmental event.

\hypertarget{data}{%
\subsection{4.1 Data}\label{data}}

We analyzed quality-screened PPI observations from MODIS and Sentinel-3 at 500 European test sites. Only finite observations satisfying the predefined usability flag were retained, without temporal interpolation. The partial years 2000 and 2025 were excluded. The composite record used MODIS observations for 2001--2018 and Sentinel-3 observations for 2019--2024, with the transition at the start of 2019. The primary seam contrast compared 2016--2018 with 2019--2021. Reference contrasts compared adjacent, non-overlapping three-year windows lying wholly within either sensor segment.

\hypertarget{s-score-calculation}{%
\subsection{4.2 S-score calculation}\label{s-score-calculation}}

A seam-blind model was fitted separately at each site by ordinary least squares. It contained a six-knot natural cubic spline for continuous time and an eight-knot cyclic cubic spline for day of year, without a sensor indicator, segment-specific intercept, or transition term.

Residual observations were assigned to the nearest of 72 nominal calendar slots---days 1, 6, 11, 16, 21, and 26 of each month---within a tolerance of three days. Each six-year contrast required at least 54 common slots spanning all 12 months, with every retained slot represented in at least two of the three years on each side. Slots were weighted equally.

The observed seam contrast was the mean post-transition residual minus the mean pre-transition residual. Matched three-year-versus-three-year reference contrasts were calculated using windows lying wholly within MODIS or wholly within Sentinel-3. The pooled reference distribution was summarized by its median and 1.4826 MAD, from which the signed S-score was calculated following Eq. (7).

A result was considered reliable when the observed contrast passed the coverage requirements, at least 10 reference contrasts were available, both record segments contributed to the reference distribution, the reference spread was at least 0.001 PPI, and all required quantities were finite. Window lengths from one to six years were examined as a descriptive sensitivity analysis, with the three-year window retained as the prespecified primary analysis. Figure 3 illustrates the fitted trajectories, residual summaries, and primary seam windows for two example sites in Spain, EU500P-177 and EU500P-188. Northern sites were not selected for illustration because missing observations limited the availability of matched contrasts.

\hypertarget{results}{%
\subsection{4.3 Results}\label{results}}

Of the 500 sites, 112 produced finite S-scores under the primary three-year design, of which 89 satisfied all reliability criteria and 23 were finite but flagged (Table 2). The remaining 388 sites did not yield a finite primary S-score. Among the 89 reliable sites, the median seam contrast was \ensuremath{-}0.006 PPI, with an interquartile range of \ensuremath{-}0.048 to +0.030. The median signed S-score was \ensuremath{-}0.020, with a range of \ensuremath{-}3.63 to +2.01; 44 sites had seam contrasts above their respective reference centers and 45 had contrasts below them. The balanced signs and near-zero median indicated little systematic departure from the reference centers across the reliable sites. These signs describe reference-centered departures rather than the raw upward or downward residual contrasts.

The median absolute S-score among reliable sites was 0.717, with an interquartile range of 0.331--1.291. Eight sites had \textbar S\textbar{} \textgreater{} 1.96, including six with \textbar S\textbar{} \textgreater{} 2 and one with \textbar S\textbar{} \textgreater{} 3. These threshold exceedances are descriptive rather than formal hypothesis-test results because the within-segment reference contrasts were finite and dependent and the sampled sites may be spatially correlated.

Reference support was strongly unbalanced between the two segments. The reliable sites had a median of 14 reference contrasts, comprising 13 from MODIS and one from Sentinel-3. The short Sentinel-3 segment therefore provided limited information about ordinary transitions within the later record, and the pooled reference distribution was dominated by MODIS contrasts.

\begin{longtable}[]{@{}
  >{\raggedright\arraybackslash}p{(\columnwidth - 2\tabcolsep) * \real{0.3958}}
  >{\raggedright\arraybackslash}p{(\columnwidth - 2\tabcolsep) * \real{0.6042}}@{}}
\caption{S-score results for the MODIS-to-Sentinel-3 vegetation record. Distribution summaries refer to the 89 sites meeting all primary reliability criteria.}\\
\toprule\noalign{}
\begin{minipage}[b]{\linewidth}\raggedright
Statistic
\end{minipage} & \begin{minipage}[b]{\linewidth}\raggedright
Result
\end{minipage} \\
\midrule\noalign{}
\endhead
\bottomrule\noalign{}
\endlastfoot
Finite primary S-scores & 112 of 500 (89 Reliable; 23 Finite but flagged) \\
Undefined under the primary design & 388 of 500 \\
Reference support & Median 14 contrasts: 13 MODIS and 1 Sentinel-3 \\
Seam contrast, \(\Delta_{\mathrm{seam}}\) & Median \ensuremath{-}0.006 PPI; IQR \ensuremath{-}0.048 to +0.030 \\
Signed S-score & Median \ensuremath{-}0.020; range \ensuremath{-}3.63 to +2.01 \\
Absolute S-score & Median 0.717; IQR 0.331--1.291 \\
Departure relative to the reference center & 44 above and 45 below their respective reference centers \\
Large standardized seams & 8/89 with \textbar S\textbar{} \textgreater{} 1.96; 6/89 with \textbar S\textbar{} \textgreater{} 2; 1/89 with \textbar S\textbar{} \textgreater{} 3 \\
\end{longtable}

Figures 3 and 4 illustrate the analysis at EU500P-177 and EU500P-188. Figure 3 shows the observations, seam-blind fitted trajectories and annual residual means, with the primary comparison windows spanning MODIS 2016--2018 and Sentinel-3 2019--2021. Figure 4(a, c) compares each observed seam contrast with matched within-segment reference contrasts. Blue and green points distinguish MODIS and Sentinel-3 references, while dark blue and red vertical lines mark the pooled reference median and observed seam contrast, respectively. The reference spread was calculated as 1.4826 \ensuremath{\times} MAD. The primary S-scores were +0.732 at EU500P-177 and +0.717 at EU500P-188, placing both observed contrasts above their respective reference centers by less than one reference spread.

The window-sensitivity analysis showed that the magnitude and sign of S varied with window length at both example sites (Figure 4b, d). Filled and open circles distinguish estimates meeting all reliability criteria from flagged estimates. Red dashed vertical lines identify the primary three-year window, and horizontal black lines indicate S = 0. The grey dashed curves show the median across sites with finite scores at each window length. Because the contributing sites may differ between window lengths, these curves describe available-case summaries rather than changes within a fixed set of sites. The site-level variation demonstrates that the three-year scores summarize continuity over the specified comparison windows and should not be interpreted as independent of window choice.

\begin{figure}[htbp]
\centering
\includegraphics[width=.85\linewidth]{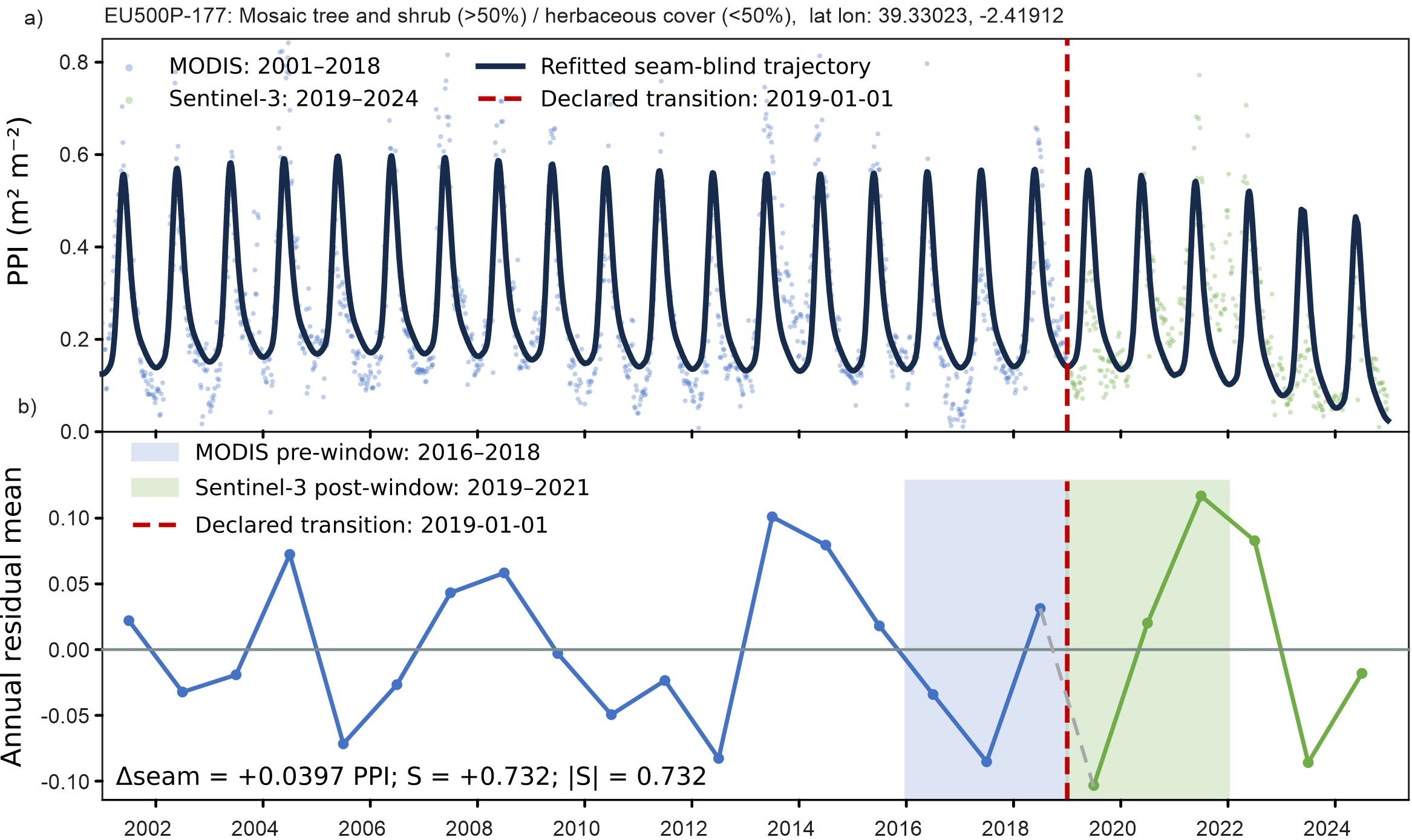}\par
\includegraphics[width=.85\linewidth]{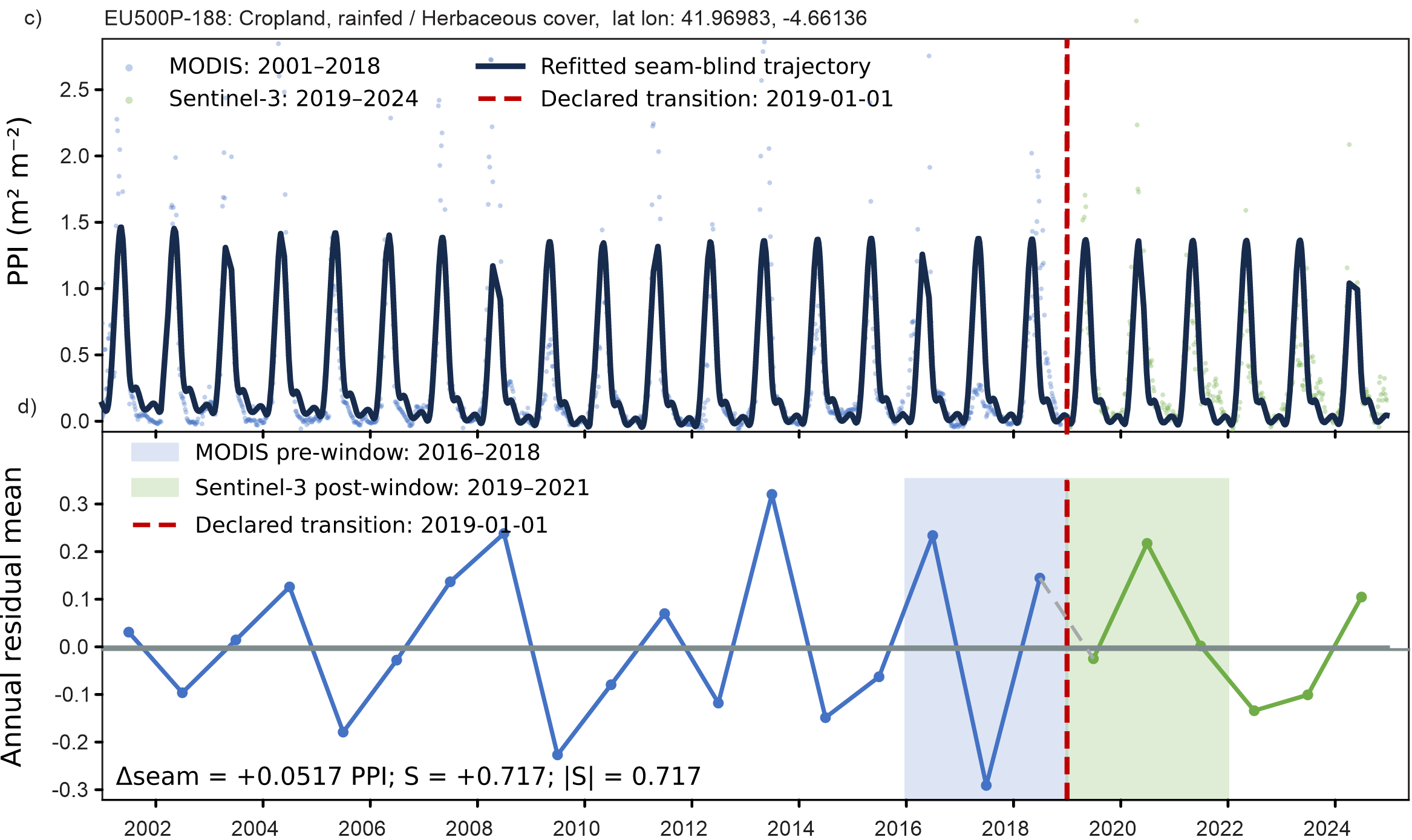}\par
\caption{\textbf{PPI records across the MODIS-to-Sentinel-3 transition at EU500P-177 (a, b) and EU500P-188 (c, d).} Upper panels show observations and seam-blind fitted trajectories; lower panels show annual residual means. Blue and green distinguish MODIS and Sentinel-3. Red dashed lines mark 1 January 2019; shading identifies the primary comparison windows, 2016--2018 and 2019--2021.}
\end{figure}

\begin{figure}[htbp]
\centering
\includegraphics[width=1\linewidth]{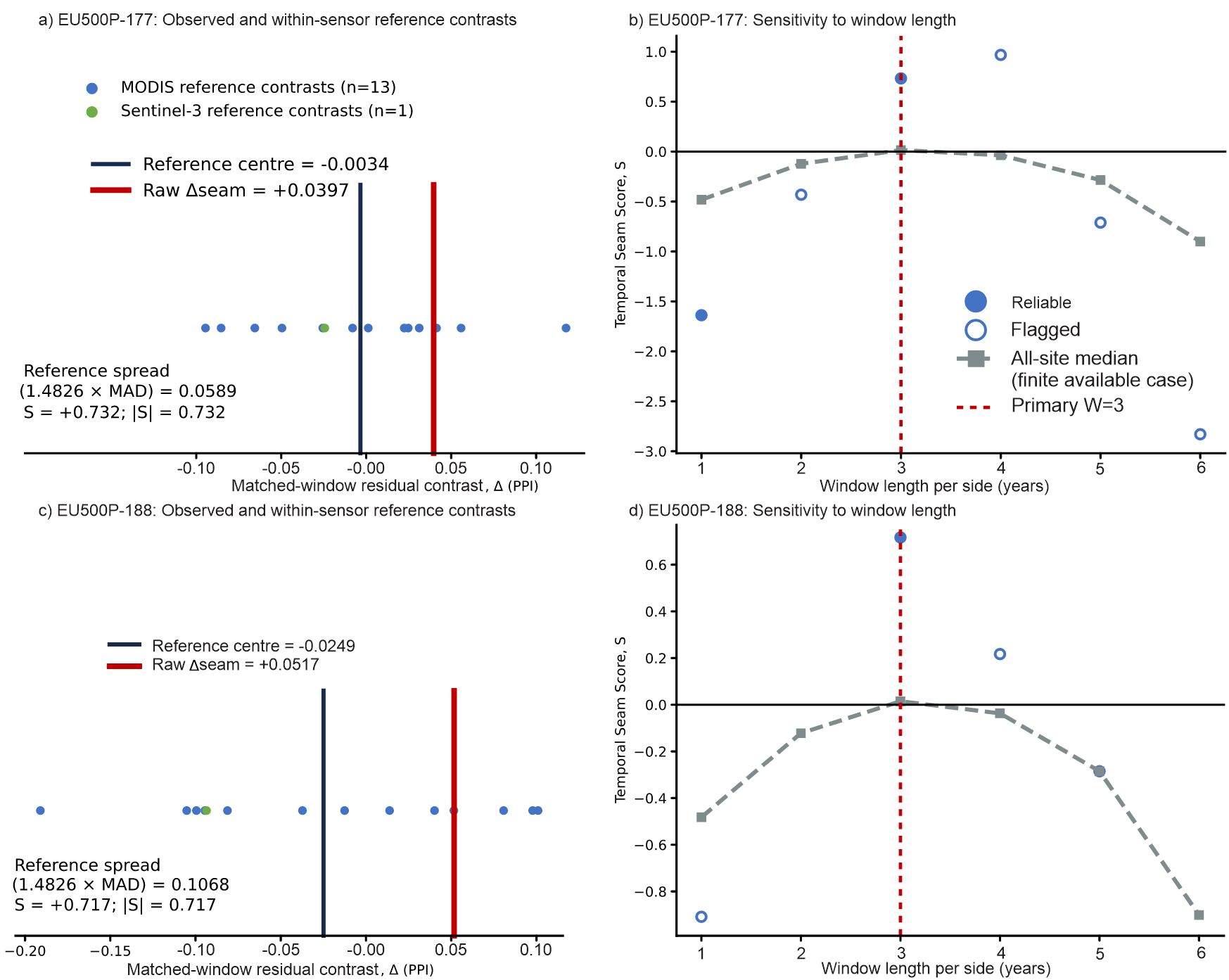}\par
\caption{\textbf{Within-segment reference contrasts and window sensitivity of the S-score at two European sites.} Panels (a, b) show EU500P-177, and panels (c, d) show EU500P-188. Panels (a, c) compare the primary seam contrast with within-segment reference contrasts; panels (b, d) show S-scores for window lengths of one to six years per side. The primary analysis compares MODIS 2016--2018 with Sentinel-3 2019--2021.}
\end{figure}

\hypertarget{discussion}{%
\section{5 Discussion}\label{discussion}}

The S-score provides a standardized and auditable assessment of continuity at a known transition in a composite time series. Its central contribution is to compare the observed seam contrast with within-segment reference contrasts over identical temporal windows. This scale matching distinguishes the S-score from visual assessment and from normalization based on individual adjacent-period changes.

The simulation quantified the relationship among seam magnitude, residual variability, and the resulting score. For example, a 20\% shift produced a median \textbar S\textbar{} of 2.58 when \({CV}_{r}\) was 0.10 but only 0.84 when \({CV}_{r}\) was 0.50. The S-score therefore measures the discontinuity relative to ordinary temporal variability rather than its percentage magnitude alone. This distinction is scientifically important: a small \textbar{}\emph{S}\textbar{} in a stable record provides stronger evidence of continuity, whereas a similarly small value in a highly variable record may indicate limited ability to resolve the seam. The raw seam contrast and reference spread are consequently essential to interpretation.

The satellite demonstration illustrated both the applicability of the method and the importance of its reliability assessment. Among the 89 sites meeting all primary criteria, the median seam contrast was \ensuremath{-}0.006 PPI and the median signed S-score was \ensuremath{-}0.020, with 44 contrasts above and 45 below their respective reference centers. These results indicate little systematic departure from the reference centers across the reliable sites, although eight sites had \textbar S\textbar{} \textgreater{} 1.96 and warranted individual investigation. More broadly, only 112 of the 500 sites produced a finite primary S-score, and 23 of these were flagged. The method therefore identifies where continuity can be assessed from the available record and where the data provide insufficient support, rather than assigning apparently precise scores to poorly constrained seams.

\hypertarget{relationship-to-existing-methods}{%
\subsection{5.1 Relationship to existing methods}\label{relationship-to-existing-methods}}

The S-score addresses a different question from change-point detection. Cumulative-sum and change-point procedures search for changes whose timing is unknown (Page, 1954; Pettitt, 1979), while multiple-break methods estimate the number and locations of structural changes (Bai \& Perron, 1998). Regression-based tests can evaluate parameter stability at a prespecified transition (Chow, 1960). These methods test for structural change, whereas the S-score expresses the magnitude and direction of departure of a known seam contrast from the center of comparable transitions within the same record.

The framework is also distinct from time-series homogenization. Homogenization methods detect and often adjust inhomogeneities using neighboring records, metadata, or statistical reference series (Venema et al., 2012). The S-score neither estimates a correction nor constructs a homogenized record. It provides a diagnostic that can be applied before harmonization to identify problematic seams or afterwards to evaluate the continuity achieved.

When simultaneous observations from two systems are available, paired agreement measures offer complementary information. Lin's concordance correlation coefficient evaluates how closely paired observations follow the one-to-one line (Lin, 1989). The S-score instead evaluates consecutive record segments and therefore remains applicable when temporal overlap is absent. Cross-sensor products such as Harmonized Landsat--Sentinel-2 demonstrate the importance of spectral, geometric, and atmospheric harmonization for constructing multi-sensor records (Claverie et al., 2018). The S-score adds a temporal diagnostic of the resulting composite record.

\hypertarget{continuity-and-attribution}{%
\subsection{5.2 Continuity and attribution}\label{continuity-and-attribution}}

The magnitude \textbar S\textbar{} describes how exceptional the observed seam contrast is relative to within-segment transitions, while the sign of S indicates the direction of departure relative to the reference center. Neither quantity identifies the cause. A large \textbar{}\emph{S}\textbar{} may arise from sensor calibration, spectral response, observation geometry, spatial resolution, sampling, quality screening, compositing, or processing changes. The same signature may result from disturbances like drought and fire, or human management like harvesting or land-cover conversion, or another environmental event coinciding with the transition. Attribution therefore requires independent evidence.

\hypertarget{extensions-to-other-seam-characteristics}{%
\subsection{5.3 Extensions to other seam characteristics}\label{extensions-to-other-seam-characteristics}}

The present S-score targets a shift in the level of aggregated residuals. A transition may instead alter trend, seasonal amplitude or phase, variance, temporal persistence, or observation availability. These characteristics require separate estimands because combining them into a single score would obscure their interpretation.

The same matched-reference principle could support component-specific extensions. For example, changes in fitted slopes, seasonal amplitudes, or residual variances across the known transition could be compared with corresponding changes within each segment. Such extensions should be named and validated separately before any multivariate seam assessment is constructed.

\hypertarget{limitations-and-reporting-requirements}{%
\subsection{5.4 Limitations and reporting requirements}\label{limitations-and-reporting-requirements}}

The S-score depends on the seam-blind model, aggregation unit, window length, and reference population. An overly flexible model may absorb a local discontinuity, whereas an insufficiently flexible model may leave gradual temporal structure in the residuals. The model and primary window should therefore be prespecified and accompanied by focused sensitivity analysis.

Reference support is especially important for short or unequal segments. In the demonstration, the reliable three-year results had a median of 13 MODIS reference contrasts but only one Sentinel-3 contrast. Pooled normalization was therefore dominated by the longer segment. Moreover, contrasts derived from overlapping windows are dependent, so their nominal count overstates the independent information available.

Missing or seasonally uneven observations can prevent construction of comparable windows. In the demonstration, 388 sites yielded no finite primary S-score; the aggregate counts alone do not identify the reasons for these undefined results. Coverage requirements should not be relaxed automatically to increase the number of finite scores. Shared or hierarchical reference distributions may improve support in spatial applications, but they require explicit assumptions about comparability and spatial dependence.

Finally, the S-score is a continuous standardized effect size rather than a universal decision rule. A small \textbar{}\emph{S}\textbar{} indicates that the seam is not exceptional relative to the estimated reference variability; it does not by itself establish radiometric or practical equivalence. Conversely, a large \textbar{}\emph{S}\textbar{} identifies an anomalous contrast but does not establish its cause. Formal flagging requires thresholds calibrated to the specific record length, window, dependence structure, missingness, and reference estimator. At minimum, applications should report \(\Delta_{\mathrm{seam}}\), signed S, \textbar S\textbar, the reference center and spread, reference counts by segment, and all reliability warnings.

\hypertarget{conclusions}{%
\section{6 Conclusions}\label{conclusions}}

The Temporal Seam Score quantifies the magnitude and direction of departure of a seam contrast at a known transition from the within-segment reference center. The sign of \(\Delta_{\mathrm{seam}}\) separately gives the raw upward or downward direction of the residual contrast. The method removes expected temporal structure, calculates a local residual contrast across the transition, and standardizes that contrast against matched within-segment references.

Simulation showed that \textbar{}\emph{S}\textbar{} is governed by seam magnitude shift relative to residual variability: the same absolute or proportional shift is readily distinguishable in a stable record but may remain unresolved in a noisy one. The MODIS-to-Sentinel-3 demonstration further showed that finite and unbalanced reference support can determine whether a site-level assessment is reliable. These findings make data sufficiency and uncertainty integral parts of continuity assessment.

The S-score provides a common framework for evaluating transitions among sensors, platforms, processing algorithms, station configurations, reanalyses, and other components of long-term records. Used together with the raw seam contrast and transparent reliability diagnostics, it can support harmonization assessment, quality assurance, and targeted investigation of discontinuities while preserving a clear separation between continuity assessment and causal attribution.

\hypertarget{code-and-data-availability}{%
\section{Code and data availability}\label{code-and-data-availability}}

This study analyzed existing MODIS and Sentinel-3 PPI time series; no new observational data were collected. Custom scripts were used for data preparation, residual modelling, calendar-matched contrast construction, and S-score calculation. These scripts are not publicly available.

\hypertarget{author-contributions}{%
\section{Author contributions}\label{author-contributions}}

Hongxiao Jin: Conceptualization, Methodology, Software, Validation, Formal analysis, Visualization, and Writing - review and editing.

\hypertarget{competing-interests}{%
\section{Competing interests}\label{competing-interests}}

The author declares no competing interests.

\hypertarget{acknowledgements}{%
\section{Acknowledgements}\label{acknowledgements}}

The simulation and demonstration were made on LUNARC high performance computation system within National Academic Infrastructure for Supercomputing in Sweden (NAISS) through an infrastructure grant awarded to Prof. Jonas Ardö by the Faculty of Science, Lund University. The demonstration data are from European Environment Agency (EEA) under a service contract to Flemish Institute for Technological Research (VITO).

\hypertarget{ai-usage}{%
\section{AI usage}\label{ai-usage}}

OpenAI GPT-6.1 Sol was used to assist with literature searches, checks of manuscript conformity with journal requirements, and improvements to language and readability. The author reviewed and verified all AI-assisted outputs and takes full responsibility for the accuracy, integrity and originality of the work.

\hypertarget{references}{%
\section{References}\label{references}}

Bai, J., \& Perron, P. (1998). Estimating and Testing Linear Models with Multiple Structural Changes. \emph{Econometrica}, \emph{66}(1), 47-78. \url{https://doi.org/10.2307/2998540}

Box, G. E. P., \& Tiao, G. C. (1975). Intervention Analysis with Applications to Economic and Environmental Problems. \emph{Journal of the American Statistical Association}, \emph{70}(349), 70-79. \url{https://doi.org/10.1080/01621459.1975.10480264}

Cai, Z., Jönsson, P., Jin, H., \& Eklundh, L. (2017). Performance of Smoothing Methods for Reconstructing NDVI Time-Series and Estimating Vegetation Phenology from MODIS Data. \emph{Remote Sensing}, \emph{9}(12), 1271. \href{http://www.mdpi.com/2072-4292/9/12/1271}{https://doi.org/10.3390/rs9121271}

Chow, G. C. (1960). Tests of Equality Between Sets of Coefficients in Two Linear Regressions. \emph{Econometrica}, \emph{28}(3), 591-605. \url{https://doi.org/10.2307/1910133}

Claverie, M., Ju, J., Masek, J. G., Dungan, J. L., Vermote, E. F., Roger, J.-C., Skakun, S. V., \& Justice, C. (2018). The Harmonized Landsat and Sentinel-2 surface reflectance data set. \emph{Remote Sensing of Environment}, \emph{219}, 145-161. \url{https://doi.org/10.1016/j.rse.2018.09.002}

Jamali, S., Jönsson, P., Eklundh, L., Ardö, J., \& Seaquist, J. (2015). Detecting changes in vegetation trends using time series segmentation. \emph{Remote Sensing of Environment}, \emph{156}(0), 182-195. \href{http://dx.doi.org/10.1016/j.rse.2014.09.010}{https://doi.org/10.1016/j.rse.2014.09.010}

Jin, H., \& Eklundh, L. (2014). A physically based vegetation index for improved monitoring of plant phenology. \emph{Remote Sensing of Environment}, \emph{152}(0), 512-525. \url{https://doi.org/10.1016/j.rse.2014.07.010}

Jin, H., Jönsson, A. M., Olsson, C., Lindström, J., Jönsson, P., \& Eklundh, L. (2019). New satellite-based estimates show significant trends in spring phenology and complex sensitivities to temperature and precipitation at northern European latitudes. \emph{International Journal of Biometeorology}, \emph{63}(6), 763--775. \url{https://doi.org/10.1007/s00484-019-01690-5}

Kennedy, R. E., Yang, Z., \& Cohen, W. B. (2010). Detecting trends in forest disturbance and recovery using yearly Landsat time series: 1. LandTrendr --- Temporal segmentation algorithms. \emph{Remote Sensing of Environment}, \emph{114}(12), 2897-2910. \url{https://doi.org/10.1016/j.rse.2010.07.008}

Lin, L. I.-K. (1989). A Concordance Correlation Coefficient to Evaluate Reproducibility. \emph{Biometrics}, \emph{45}(1), 255-268. \url{https://doi.org/10.2307/2532051}

Page, E. S. (1954). Continuous inspection schemes. \emph{Biometrika}, \emph{41}(1-2), 100-115. \url{https://doi.org/10.1093/biomet/41.1-2.100}

Pettitt, A. N. (1979). A Non-Parametric Approach to the Change-Point Problem. \emph{Journal of the Royal Statistical Society. Series C (Applied Statistics)}, \emph{28}(2), 126-135. \url{https://doi.org/10.2307/2346729}

Venema, V. K. C., Mestre, O., Aguilar, E., Auer, I., Guijarro, J. A., Domonkos, P., Vertacnik, G., Szentimrey, T., Stepanek, P., Zahradnicek, P., Viarre, J., Müller-Westermeier, G., Lakatos, M., Williams, C. N., Menne, M. J., Lindau, R., Rasol, D., Rustemeier, E., Kolokythas, K., . . . Brandsma, T. (2012). Benchmarking homogenization algorithms for monthly data. \emph{Clim. Past}, \emph{8}(1), 89-115. \url{https://doi.org/10.5194/cp-8-89-2012}

Verbesselt, J., Hyndman, R., Newnham, G., \& Culvenor, D. (2010). Detecting trend and seasonal changes in satellite image time series. \emph{Remote Sensing of Environment}, \emph{114}(1), 106-115. \url{https://doi.org/10.1016/j.rse.2009.08.014}

Zhu, Z., \& Woodcock, C. E. (2014). Continuous change detection and classification of land cover using all available Landsat data. \emph{Remote Sensing of Environment}, \emph{144}, 152-171. \url{https://doi.org/10.1016/j.rse.2014.01.011}

\end{document}